\documentclass[journal,10pt,twoside]{IEEEtran}%

\usepackage{cite}

\ifCLASSINFOpdf
\else
\fi
\usepackage[cmex10]{amsmath}
\usepackage{algorithmic}

\usepackage{array}

\usepackage[tight,footnotesize]{subfigure}
\usepackage{url}

\usepackage{color}

\usepackage{amssymb}
\usepackage{amsthm}
\theoremstyle{plain}

\theoremstyle{definition}
\newtheorem{defn}{Definition}
\newtheorem{note}{Note}
\theoremstyle{remark}

\input xy
\xyoption{all}
\xyoption{color}

\usepackage{wasysym,multirow,pifont,hyperref,algorithm2e,mathtools}
\usepackage{xcolor}
\usepackage{graphicx}
\usepackage{stfloats}
\usepackage{float}

\IEEEiedlistdecl

\allowdisplaybreaks[4]
\newcounter{MYtempeqncnt}

\begin{document}
\title{Controlling the Error Floor in LDPC Decoding}

\author{Shuai~Zhang and Christian~Schlegel,~\IEEEmembership{Fellow,~IEEE}
\thanks{S.~Zhang is with the High Capacity Digital Communications
Laboratory (HCDC), University of Alberta, Edmonton AB, T6G 2V4, Canada (e-mail: szhang4@ualberta.ca).

C.~Schlegel is with the NSERC Ultra Marine Digital Communications Center, Dalhousie University,
Halifax NS, B3H 4R2, Canada (e-mail: Christian.Schlegel@Dal.ca).

Parts of this material was previously presented at the 49th Allerton
Conference on Communications,
Control, and Computing, and at the 2012 Information Theory and Applications workshop in San Diego.}
}
%
%

\markboth{Submitted to IEEE Transactions on Communications}%
{Zhang \MakeLowercase{\textit{et al.}}: Controlling the Error Floor in LDPC Decoding}
%

\IEEEpubid{0000--0000/00\$00.00~\copyright~2012 IEEE}


\maketitle

\begin{abstract}
The error floor of LDPC is revisited as an effect of dynamic message
behavior in the so-called absorption sets of the code. It is shown
that if the signal growth in the absorption sets is properly
balanced by the growth of set-external messages, the error floor can
be lowered to essentially arbitrarily low levels. Importance
sampling techniques are discussed and used to verify the analysis,
as well as to discuss the impact of iterations and message
quantization on the code performance in the ultra-low BER (error
floor) regime.
\end{abstract}


%
\IEEEpeerreviewmaketitle

\section{Introduction}
%
%
%
%
\IEEEPARstart{L}{ow}-density parity-check (LDPC) codes are a class
of linear block codes \cite{gallager} which have enjoyed intense
theoretical and practical attention due to their excellent
performance and decoding efficiency. Their parity-check matrices are
given by sparse binary $(0,1)$ matrices which enable very efficient
$O(n)$-complexity decoder implementations, where $n$ is the code
length. The error rate of LDPC codes decreases rapidly as the
signal-to-noise ratio (SNR) increases, and comes very close to
maximum-likelihood (ML) decoding error performance, which is
determined by  the distance spectrum of the code.

However, when utilizing sub-optimal $O(n)$ iterative decoding algorithm, such as message passing (MP)
or linear programming, a marked increase of error rates at high SNRs tends to appear with respect to
optimal decoding. The error curve assumes the shape of an error floor with a very slow decrease with SNR. This error floor is
caused by inherent structural weaknesses in the code's interconnect network, which cause long
latencies in the iterative decoder or outright lock-up in error patterns associated with these
weaknesses. These decoding failures are very important for low-error-rate applications such as
cable modems and optical transmission systems. They were initially studied in \cite{wiberg, frey, forney},
and called {\em trapping sets} in \cite{Rich04}. Such trapping sets are dependent not only
on the code but also on the channel where the code is used, as well as the specific decoding algorithm.

For example, trapping sets on the binary erasure channel are known as
{\em stopping sets} \cite{ChanProiTelaRichUrba02}, whereas the dominant trapping sets of LDPC
codes on the Gaussian channel are the {\em absorption sets} \cite{Zhanetal06}. Absorption sets
are also the sets implicated in the failure mechanism of message-passing decoder for
binary symmetric channels.

Due to the importance of error floors in low-BER applications, recent efforts have focused on understanding
the dynamics of absorption sets \cite{Rich04, oscar, itw09, Zhanetal2008,  cs09, backtracking11, allerton11, ButlerAllerton11, pascal}, and several modifications of the decoding algorithm have been studied to
lower the error floor, specifically targeting the absorption sets \cite{graphcover08,
Zhanetal08, backtracking11, allerton11}. As we discuss in this paper, the onset of the error floor on
Gaussian channels is very strongly related to the behavior of the algorithm on binary symmetric channels, and
that the dynamics of the absorption sets fully explain why and when the error floor becomes dominant
in a given code.

In our earlier work \cite{itw09, cs09, allerton11} we developed a
linear model to analyze dynamics of and probabilities for the error
floor for specific codes. We identified and enumerated the
{\em minimal} absorption sets, which dominate the decoding
performance in the error floor region for the IEEE 802.3an and the
Tanner $[155,64,20]$ codes, and derived closed-form formulas that
approximate the probability that an absorption set falls in error.
\IEEEpubidadjcol

In this paper, we first refine the formula for the error floor probability \cite{itw09,
cs09, allerton11}, and use it to show that the error floor critically depends on the
limits, aka thresholds, which are utilized to represent the internal log-likelihood messages in the
decoder. We show that the growth rate of the error patterns in the absorption set can be balanced
by the growth of the LLRs external to the set, if a sufficient dynamic range for these messages is available.
We quantify this effect and compare it to importance sampling guided simulations for verification.
We verify the hypothesis that, quite unlike in the case of binary erasure or binary symmetric channels,
the error floor phenomenon for LDPC codes on AWGN channels  is essentially an ``artifact'' of inexact
decoding, and not an inherent weakness of the code.
We then quantify the effect of the number of iterations and show agreement of our refined equations
with simulated error floor rates for both finite iterations and thresholds. Finally, we examine the
impact of message quantization on the error floor and thus the required computational complexity
to attain a given level of performance in the ultra-low BER regime of the code.

We wish to note that while sensitivity of the error floor to message
thresholding has been observed heuristically by a number of authors,
the view presented in this paper appears to have been developed
independently by Butler and Siegel \cite{ButlerAllerton11} as well
as the authors \cite{allerton11,itw09}.

This paper is organized as follows: In Section~\ref{bg}, LDPC codes,
iterative decoding, and absorption sets are reviewed, and the two
example LDPC codes used in our studies are introduced. In
Section~\ref{pr}, our linear algebraic approach to evaluate the
error rate is described as  a two-step procedure. The first step is
to identify the dominant error patterns and the second is performing
the analysis targeting these error patterns. Based on the insights
provided by the analytical approach, Section~\ref{boosting} studies
different decoder settings for both the length-$155$ Tanner and the
IEEE 802.3an LDPC codes. The computational complexity and importance
sampling are discussed in Section~\ref{secv}. Finally, the
conclusion will be given in Section~\ref{end}.

\section{Background}\label{bg}

In this section, we first review the main features of LDPC codes and the most popular iterative decoding algorithms. Two LDPC codes will be introduced, as well. Then we apply the linear analysis technique of
\cite{itw09} to study their error floors in the following sections.

\subsection{LDPC Codes}

\begin{figure}[!t]
\begin{center}
\setlength{\unitlength}{0.75mm}
\begin{picture}(180,60)
\put(-4,-20){\includegraphics[scale=0.57]{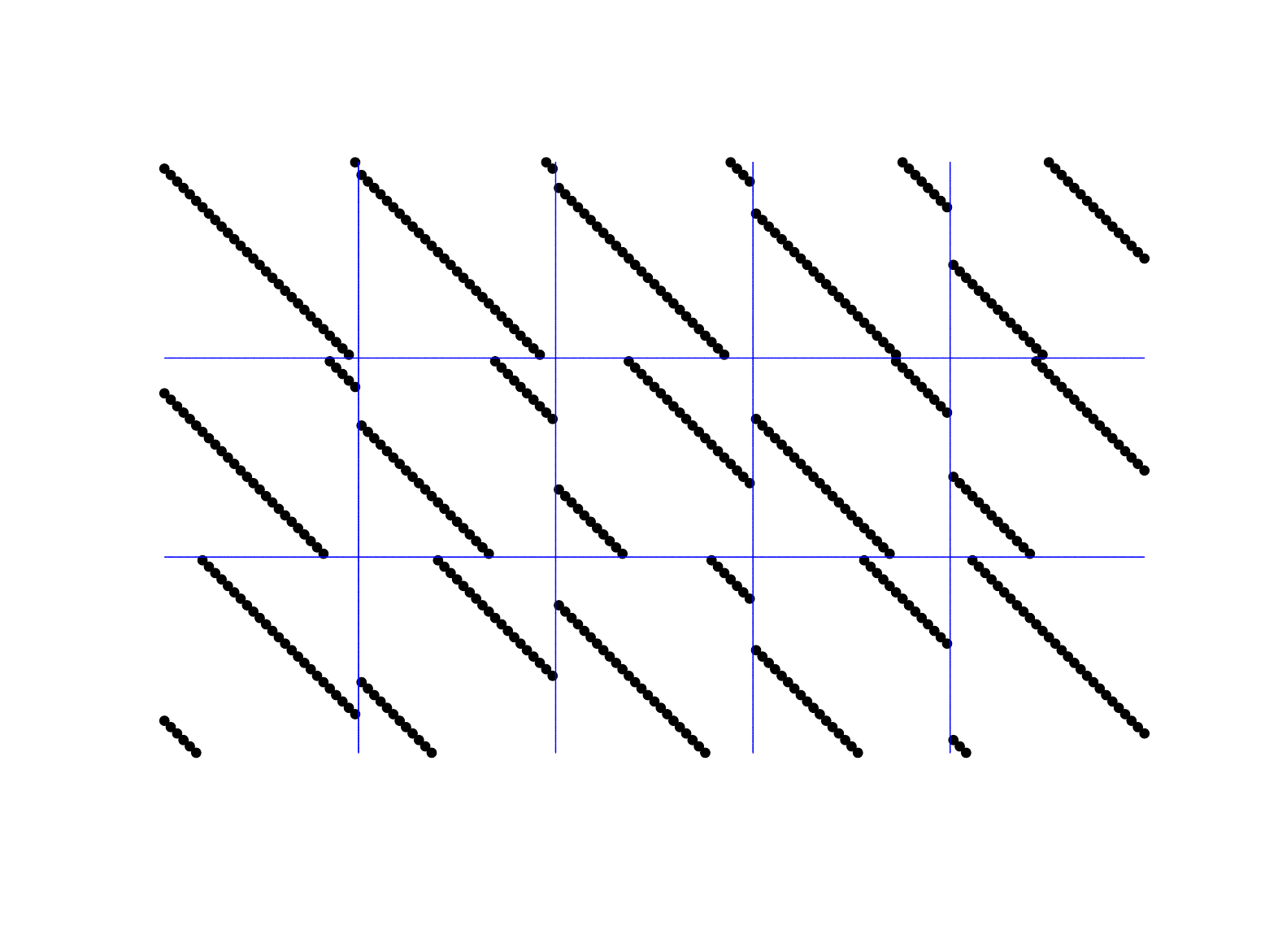}}
\end{picture}
\caption{The parity-check matrix of the Tanner $[155, 64, 20]$ regular $(3,5)$ LDPC code.}\label{tanner155bin2}
\end{center}
\end{figure}

An LDPC code is associated with a sparse parity-check matrix,
denoted by $\mathbf{H}_{m\times n}$, where every column represents a
code bit, and every row constitutes a parity-check equation. The
code length is $n$, and the number of information bits $k\geq n-m$,
since $\mathbf{H}$ is not necessarily full rank. If every column and
every row of $\mathbf{H}$ has the same number of non-zero elements,
then we have a ``regular'' LDPC code, otherwise the code is
irregular. A regular LDPC code can also be represented by $(d_v,
d_c)$, where $d_v$ and $d_c$ are the Hamming weights of each column
and each row, respectively, together with an interleaver.
\figurename~\ref{tanner155bin2} shows the parity-check matrix of a
$(3,5)$ regular LDPC code, where non-zero elements are plotted as
solid dots and zeros are left blank.

Following Tanner, an LDPC code can be represented by a bipartite
graph, called a Tanner graph. Let check nodes denote the rows of the
parity-check matrix, and variable nodes denote the columns of the
matrix. Then every non-zero entry in $\mathbf{H}$ indicates a
connection between these two disjoint sets.

\subsubsection{Tanner $[155, 64, 20]$ Regular $(3,5)$ LDPC Code}

In \cite{tanner2001} and \cite{tanner2004}, Tanner introduced a
class of regular LDPC codes composed of blocks of circulant permutation
matrices. The parity-check matrix
$\mathbf{H}$ of these codes consists of a $d_v \times d_c$ array of
such circulant permutation matrices $\mathbf{I}_k$, where each
$\mathbf{I}_k$ denotes a $p \times p$ identity matrix with its
rows shifted cyclically to the left by $k$ positions.
\begin{equation}\label{eq:tannerH}
\mathbf{H}=\begin{bmatrix*}[l]
\mathbf{I}_{0,0} & \mathbf{I}_{0,1} & \cdots & \mathbf{I}_{0,d_c-1}\\
\mathbf{I}_{1,0} & \mathbf{I}_{1,1} & \cdots & \mathbf{I}_{1,d_c-1}\\
\vdots& \vdots& \ddots& \vdots\\
\mathbf{I}_{d_v-1,0} & \mathbf{I}_{d_v-1,1} & \cdots & \mathbf{I}_{d_v-1,d_c-1}\\
\end{bmatrix*}_{d_v p \times d_c p}.
\end{equation}
Immediately, we can tell that each row of \eqref{eq:tannerH} has
exactly $d_c$, and each column $d_v$ non-zero elements, making the
code a regular $(d_v,d_c)$ LDPC code. Its length is $d_c p$ and the
number of check nodes is $d_v p$. Its ``design'' code rate is
$1-d_v/d_c$, however the actual rate $R$ will be slightly higher in
that within every row block, all $p$ rows add to an all-one vector
and at least $d_v-1$ rows are linearly dependent.

The dimension of the base identity matrix $p$ was originally
designed to be prime to eliminate $4$-cycles. In \cite{tanner2004},
$p$ was extended to non-primes, where the girth, the shortest cycle
in the Tanner graph and denoted by $g$, can be as short as $4$. It
can be shown that $g$ is upper bounded by $12$, no matter how large
$n$ is \cite{tanner2001, tannergirth04}. The minimum
distance of a Tanner code is bounded by $d_{\min}\leq (d_v+1)!$
\cite{MacKayHighRate98}. When $d_v$ is small, this upper bound can
be met by carefully selecting the parameters \cite{MacKayHighRate98,
tanner2004}. Tanner codes come with relatively large girth and/or
minimum distance, and provide good error floor performance. However,
low-weight absorption sets are not eliminated by this construction.

Tanner presented a $[155, 64, 20], (3,5)$ regular LDPC code in the ``Recent Results'' session of the IEEE International Symposium on Information Theory (ISIT) in 2000. Its block structure parity-check matrix is given by
\begin{equation}
\mathbf{H}=\begin{bmatrix*}[l]
\mathbf{I}_{1} &\mathbf{I}_{2} & \mathbf{I}_{4} & \mathbf{I}_{8} & \mathbf{I}_{16} \\
\mathbf{I}_{5} & \mathbf{I}_{10} & \mathbf{I}_{20} & \mathbf{I}_{9} & \mathbf{I}_{18} \\
\mathbf{I}_{25} & \mathbf{I}_{19} & \mathbf{I}_{7} & \mathbf{I}_{14} & \mathbf{I}_{28} \\
\end{bmatrix*}_{93\times 155},
\end{equation}
where each $\mathbf{I}_{x}$ is derived by shifting the rows of a $31
\times 31$ identity matrix cyclically to the left by $x$ positions.
Its binary structure is shown in \figurename~\ref{tanner155bin2}.
This Tanner code has rate $R\approx 0.4129$ and is equipped with a
large girth $g=8$ and $d_{\min}=20$ \cite{tanner2001}. We will use
this code as one of the examples to illustrate our error floor
analysis.

\subsubsection{IEEE 802.3an $[2048, 1723]$ Regular $(6,32)$ LDPC Code}

The IEEE 802.3an RS-based low-density parity-check code belongs to a special class of binary linear block codes, constructed in \cite{DjuXuAbdLin03, ShuLin05}. Its parity-check matrix is also comprised of blocks of permutation matrices:
\begin{equation}
\mathbf{H}=\begin{bmatrix}
\sigma_{1,1} &\sigma_{1,2} &\cdots &\sigma_{1,32} \\
\sigma_{2,1} &\sigma_{2,2} &\cdots &\sigma_{2,32} \\
\vdots& \vdots& \ddots& \vdots\\
\sigma_{6,1} &\sigma_{6,2} &\cdots &\sigma_{6,32} \\
\end{bmatrix}_{384\times 2048},
\end{equation}
where each $\sigma_{i,j}$ is a $64\times
64$ permutation matrix.

The design code rate is $1-d_v/d_c=0.8125$, whereas the actual code
rate is slightly higher with $R=1723/2048\approx0.8413$. The Tanner
graph representing this code is $4$-cycle free and the minimal cycle
length is $g=6$. We believe the minimum distance of this code is
$14$ (it is either $12$ or $14$ \cite{cs09}).

\subsection{Message-Passing Iterative Decoding Algorithms}

Although ML decoding is optimal, it is too complex to implement. On the other hand, $O(n)$-iterative
decoding performs extremely well. The idea of iterative decoding is that each check node combines all the information sent to it and returns a likelihood/suggestion back to the variable nodes connected to it. The
variable nodes can either make a decision of the incoming messages, for example, a majority vote, or
combine them up and send them back out to the check nodes for another iteration cycle. The process will stop when all check nodes are satisfied, i.e., a valid codeword is encountered, or the decoder reaches a maximum number of iterations allowed.

Decoding is monotonic in that the more iterations the decoder executes, the better the performance. This is
strictly true for cycle-free codes. When the codes have cycles, the messages passing along the edges
become dependent after a few iterations, which degrades the final decoding performance.
Therefore, constructing a code with large girth is one way to guarantee good performance.
In this paper a standard log-likelihood message-passing (MP) iterative decoding algorithm will be
used as benchmark.

\subsection{LDPC Absorption Sets}

The Gaussian channel differs from the binary symmetric channel and
the binary erasure channel in that the dynamics of the error
behavior of the LDPC decoder is more complicated. Richardson
observed that the failure mechanism in the error floor of Gaussian
channels was caused by what he called {\em trapping sets}
\cite{Rich04}. But fully classifying trapping sets is a largely
unsolved, and perhaps unsolvable, problem. Nonetheless, subsequent
investigations noted that a special class of trapping sets, called
{\em absorption sets}, dominates the error floor region
\cite{Zhanetal06, Zhanetal2008}. Furthermore, minimal absorption
sets (see later) play a critical role in the error floor.

\begin{defn}\label{asdef}
An absorption set is a set of variable nodes such that the majority of the
neighboring (connected) check nodes of each variable node in the set are
connected to the set an even number of times.
\end{defn}

\begin{note}\label{note1}
The importance of absorption sets can easily be appreciated by
realizing that Gallager's original bit flipping decoding algorithm
\cite{gallager} will fail to correct an error pattern that falls
onto an absorption set, and that such an error will therefore
persist, even if the iteration count goes to infinity.
\end{note}

Let an ordered pair $(a,b)$ denote an absorption set where $a$ is
size of the set and $b$ is the extrinsic message degree, i.e., the
cardinality of the set (no repetition allowed) of the neighboring
check nodes that are connected to the set an {\em odd} number of
times (usually once). The belief that ``smaller''  absorption sets causes more
severe effects on the error floor \cite{Rich04,elementaryas05} is
supported by the analytical error floor equations derived in
\cite{itw09, cs09}. This is analogous to the fact that lower weight
codewords have more impact on the error rate than higher weight
codewords.


\section{Error Floor Estimation}\label{pr}

Our linear algebraic estimation process  proceeds in two steps.
First, the dominant absorption set topologies along with their
multiplicities have to be identified for a given code. These are
required to compute the dynamics of the sets in step two.

\subsection{Absorption Set Identification}

There exist several papers on absorption set enumeration which make
use of the topological features of the sets as well as algebraic
properties used in the construction of the codes, \cite{itw09, cs09,
5454095, Zhanetal2008}. The search for absorption sets can be either
algebraic or simulation-based, where the latter typically produces lower bounds
on the multiplicities and set varieties. We have modified existing
techniques and integrated them into our search method, in order to
exhaustively enumerate the topologies and multiplicities in a
systematic fashion for the more dominant sets up to a certain size,
limited by computational complexity.

In the following, we list topological properties of the absorption sets of the Tanner
$[155, 64, 20]$ and the IEEE802.3an $[2048, 1723]$ LDPC codes, respectively.

\subsubsection{Tanner $[155, 64, 20], (3, 5)$ LDPC Code}

Following the enumerating methods in \cite{itw09, cs09}, we list the
first few absorption sets of the Tanner $[155, 64, 20], (3, 5)$ LDPC
code in \tablename~\ref{table155as} \cite{allerton11}.

\begin{table}[!t]
\caption{First few absorption sets of Tanner $[155, 64, 20], (3, 5)$
code.} \label{table155as}
\begin{center}
\begin{tabular}{|c|c|c|r|c|}\hline
$a$ & $b$ & \textbf{Existence} & \textbf{Multiplicity} &
\textbf{Gain} ($\mu_{\max}$)\\\hline\hline $<4$ & & No &&\\\hline $4$
& $4$ & Yes & $465$ & $1$ \\\hline \multirow{3}{*}{$5$} & $1$ & No &
& \\\cline{2-5}
 & $3$ & \multirow{2}{*}{Yes} & $155$ & \\\cline{2-2}\cline{4-5}
 & $5$ &  & $3,720$ & \\\hline
\multirow{3}{*}{$6$} & $2$ & No & & \\\cline{2-5}
 & $4$ & \multirow{2}{*}{Yes} & $930$ & \\\cline{2-2}\cline{4-5}
 & $6$ &  & $22,630$ & $1$ \\\hline
\multirow{4}{*}{$7$}  & $1$ & No &  &  \\\cline{2-5}
 & $3$ & \multirow{7}{*}{Yes} & $930$ &  \\\cline{2-2}\cline{4-5}
 & $5$ &  & $16,275$ & \\\cline{2-2}\cline{4-5}
 & $7$ &  & $140,430$ & $1$ \\\cline{1-2}\cline{4-5}
\multirow{4}{*}{$8$} & $2$ & &$465$& $1.7870$
\\\cline{2-2}\cline{4-5}
 & $4$ & & $5,115$ &\\\cline{2-2}\cline{4-5}
 & $6$ &  & $196,540 $ & \\\cline{2-2}\cline{4-5}
 & $8$ & & $823,515$ & $1$ \\ \hline
\end{tabular}
\end{center}
\end{table}%

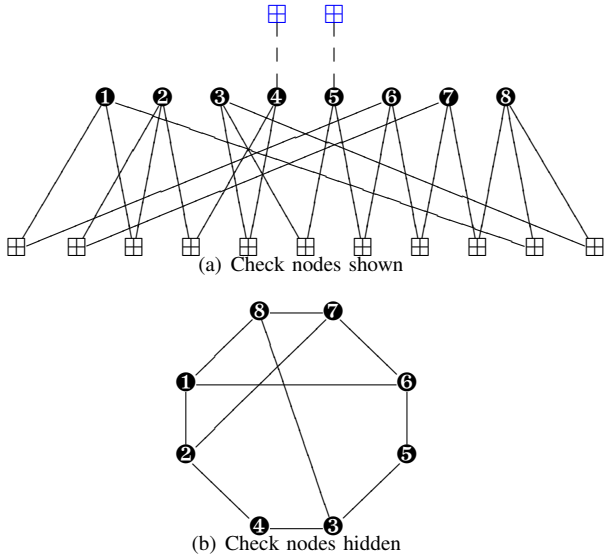
\begin{figure}[!t]
\centering \subfigure[Check nodes shown] 
{ \label{fig15582:sub:a} $\xymatrix@M=0pt@W=0pt@R=24.4pt@C=3pt {
&&&&&&&&&\color{blue}\boxplus\ar@{--}[d]&&\color{blue}\boxplus\ar@{--}[d]&&&&&&&&&\\
&&& \text{\ding{182}}&& \text{\ding{183}}&& \text{\ding{184}}&& \text{\ding{185}}&& \text{\ding{186}}&& \text{\ding{187}}&& \text{\ding{188}}&& \text{\ding{189}}&&&\\
&&&&&&&&&&&&&&&&&&&&\\
\boxplus\ar@{-}[uurrr]\ar@{-}[uurrrrrrrrrrrrr]&\hspace{9pt}&
\boxplus\ar@{-}[uurrr]\ar@{-}[uurrrrrrrrrrrrr]&&
\boxplus\ar@{-}[uur]\ar@{-}[uul]&&
\boxplus\ar@{-}[uurrr]\ar@{-}[uul]&&
\boxplus\ar@{-}[uur]\ar@{-}[uul]&&
\boxplus\ar@{-}[uur]\ar@{-}[uulll]&&
\boxplus\ar@{-}[uur]\ar@{-}[uul]&&
\boxplus\ar@{-}[uur]\ar@{-}[uul]&&
\boxplus\ar@{-}[uur]\ar@{-}[uul]&&
\boxplus\ar@{-}[uulllllllllllllll]\ar@{-}[uul]&\hspace{9pt}&
\boxplus\ar@{-}[uulllllllllllll]\ar@{-}[uulll]
 }$}\\
\subfigure[Check nodes hidden] 
{
    \label{fig15582:sub:b}
$$\xymatrix@M=0pt@W=0pt@R=20pt@C=20pt
{
& \text{\ding{189}}\ar@{-}[r]\ar@{-}[dl]\ar@{-}[dddr]& \text{\ding{188}} \ar@{-}[dr]\ar@{-}[ddll]&\\
\text{\ding{182}} \ar@{-}[d]\ar@{-}[rrr]&&& \text{\ding{187}}\ar@{-}[d]\\
\text{\ding{183}}\ar@{-}[dr]&&& \text{\ding{186}}\ar@{-}[dl]\\
& \text{\ding{185}}\ar@{-}[r]& \text{\ding{184}}&\\
} $$ } \caption {The topology of the $(8,2)$ absorption set of the Tanner $[155,64,20]$ regular $(3,5)$ LDPC code (not all
check node connections shown).}
\label{fig15582:sub} 
\end{figure}

Although the $(8,2)$ set is not the smallest absorption set in terms of $a$, it does have a
small number of unsatisfied check nodes, which makes it dominant.
\figurename~\ref{fig15582:sub} shows the subgraph induced by the $(8,2)$ absorption set.

We observe from \figurename~\ref{fig15582:sub:b} that the $(8,2)$
absorption set consists of cycles of length $8, 10, 12, 14$ and
$16$. In addition, it is also an extension set of lower weight  absorption
sets. As a matter of fact, all $(4,4)$, $(5,3)$, $(6,4)$ and $(7,3)$
sets are contained in $(8,2)$ sets. And $50\%$ of $(5,5)$, $4.1\%$
of $(6,6)$, $8.6\%$ of $(7,5)$ sets are contained in $(8,2)$ sets,
respectively. We found that this $(8,2)$ set is the dominant
absorption/trapping set of this Tanner code on the Gaussian channel
under iterative MP decoding.

Actually, \figurename~\ref{fig15582:sub:a} looks very much like a codeword in
that, if all those eight variable nodes are erroneous and the rest of the variables
in the code are correct, then all but two check nodes are unsatisfied.

\begin{figure*}[!t]
\setcounter{MYtempeqncnt}{\value{equation}} \setcounter{equation}{4}
\begin{eqnarray}
\boldsymbol{\lambda} &=& \big[\underbrace{\lambda_1, \lambda_1,
\lambda_1,}_{\text{node \ding{182}}} \underbrace{\lambda_2,
\lambda_2, \lambda_2,}_{\text{node \ding{183}}}
\underbrace{\lambda_3, \lambda_3, \lambda_3,}_{\text{node
\ding{184}}} \underbrace{\lambda_4, \lambda_4,}_{\text{node
\ding{185}}} \underbrace{\lambda_5, \lambda_5,}_{\text{node
\ding{186}}}
\underbrace{\lambda_6,\lambda_6,\lambda_6,}_{\text{node \ding{187}}} \underbrace{\lambda_7,\lambda_7,\lambda_7,}_{\text{node \ding{188}}} \underbrace{\lambda_8,\lambda_8,\lambda_8}_{\text{node \ding{189}}}\big]^\mathrm{T} \label{mlambda}\\
\boldsymbol{\lambda}^{(\text{ex})}_i &=& \Big[
\underbrace{0,0,0,}_{\text{node \ding{182}}}
\underbrace{0,0,0,}_{\text{node \ding{183}}}
\underbrace{0,0,0,}_{\text{node \ding{184}}}
\underbrace{\lambda^{(\text{ex})}_{i4},
\lambda^{(\text{ex})}_{i4},}_{\text{node \ding{185}}}
 \underbrace{\lambda^{(\text{ex})}_{i5}, \lambda^{(\text{ex})}_{i5},}_{\text{node \ding{186}}} \underbrace{0,0,0,}_{\text{node \ding{187}}} \underbrace{0,0,0,}_{\text{node \ding{188}}} \underbrace{0,0,0}_{\text{node \ding{189}}}\Big]^\mathrm{T} \label{mlambdaex}
\end{eqnarray}
\setcounter{equation}{\value{MYtempeqncnt}}\hrulefill\vspace*{4pt}
\end{figure*}

\subsubsection{IEEE 802.3an $[2048, 1723]$ LDPC Code}

Absorption sets up to size $a=10$ are listed in Table~\ref{table:8023ab}
\cite{cs09}.

\begin{table}[!t] 
\caption{The first few absorption sets of the IEEE 802.3an $[2048, 1723]$, $(6,32)$ LDPC code.}
\label{table:8023ab}
\centering
\begin{tabular}{|c|c|c|r|c|}\hline
$a$ & $b$ & \textbf{Existence} & \textbf{Multiplicity} &
\textbf{Gain} ($\mu_{\max}$)\\\hline\hline $<5$ & & No &&\\\hline $5$ & $10$ & No
&& \\\hline \multirow{4}{*}{$6$} & $6$ & \multirow{4}{*}{No}
&&\\\cline{2-2}
 & $8$ & &&\\\cline{2-2}
 & $10$ & &&\\\cline{2-2}
 & $12$ & &&\\ \hline
 \multirow{8}{*}{$7$} & $0$ & \multirow{6}{*}{No}&&\\\cline{2-2}
 & $2$ &&& \\\cline{2-2}
 & $4$ &&& \\\cline{2-2}
 & $6$ &&& \\\cline{2-2}
 & $8$ &&& \\\cline{2-2}
 & $10$ &&& \\\cline{2-5}
 & $12$ & \multirow{2}{*}{Yes} & $65,472$& $3.29$\\ \cline{2-2}\cline{4-5}
 & $14$ &  & $14,720$ & $3$ \\ \hline
 \multirow{9}{*}{$8$} & $0$ & \multirow{4}{*}{No}&&\\\cline{2-2}
 & $2$ &&& \\\cline{2-2}
 & $4$ &&& \\\cline{2-2}
 & $6$ &&& \\\cline{2-5}
 & $8$ & Yes & $14,272$ & $4$\\\cline{2-5}
 & $10$ & No &&\\\cline{2-5}
 & $12$ & \multirow{3}{*}{Yes} & $44,416$ &$3.5$\\\cline{2-2}\cline{4-5}
 & $14$ &  & $88,896$ &$3.25$\\\cline{2-2}\cline{4-5}
 & $16$ &  & $661,824$ &$3$\\ \hline
\multirow{8}{*}{$9$} & $0$ & \multirow{6}{*}{No} &&\\ \cline{2-2}
  & $2$ &  &&\\ \cline{2-2}
  & $4$ &  &&\\ \cline{2-2}
  & $6$ &  &&\\ \cline{2-2}
  & $8$ &  &&\\ \cline{2-2}
  & $10$ &  &&\\ \cline{2-5}
  & $12$ & \multirow{4}{*}{Yes} & \multirow{4}{*}{?} &$3.67$\\\cline{2-2}\cline{5-5}
  & $14$ &  & &$3.44$\\\cline{2-2}\cline{5-5}
  & $16$ &  &&$3.22$\\ \cline{2-2}\cline{5-5}
  & $18$ &  &&$3$\\ \hline
  \multirow{11}{*}{$10$} & $0$ & \multirow{4}{*}{No} &&\\ \cline{2-2}
    & $2$ &  &&\\ \cline{2-2}
    & $4$ &  &&\\ \cline{2-2}
    & $6$ &  &&\\ \cline{2-5}
  & $8$ & ? &&\\ \cline{2-5}
  & $10$ & \multirow{6}{*}{Yes} & $>192$ & $4$\\ \cline{2-2}\cline{4-5}
 & $12$ &  &\multirow{5}{*}{?}&$3.8$\\\cline{2-2}\cline{5-5}
 & $14$ &  &  &$3.6$\\\cline{2-2}\cline{5-5}
 & $16$ & &&$3.4$\\\cline{2-2}\cline{5-5}
 & $18$ &  &&$3.2$\\\cline{2-2}\cline{5-5}
 & $20$ &  &&$3$\\ \hline
\end{tabular}
\end{table}

\begin{figure}[!t]
\centering \subfigure[Check nodes shown] 
{
    \label{fig:ieee88:a}
$\xymatrix@M=0pt@W=0pt@R=24.4pt@C=-2pt
{
&&&&& \color{blue}\boxplus\ar@{--}[d]&&&& \color{blue}\boxplus\ar@{--}[d]&&&& \color{blue}\boxplus\ar@{--}[d]&&&& \color{blue}\boxplus\ar@{--}[d]&&&& \color{blue}\boxplus\ar@{--}[d]&&&& \color{blue}\boxplus\ar@{--}[d]&&&& \color{blue}\boxplus\ar@{--}[d]&&&& \color{blue}\boxplus\ar@{--}[d]&&&&& \\
&&&&& \CIRCLE\ar@{-}[ddlllll]\ar@{-}[ddlll]\ar@{-}[ddl]\ar@{-}[ddr]\ar@{-}[ddrrr]&&&& \CIRCLE\ar@{-}[ddr]\ar@{-}[ddrrr]\ar@{-}[ddrrrrr]\ar@{-}[ddrrrrrrr]\ar@{-}[ddrrrrrrrrrrrrrrrrrrrrr]&&&& \CIRCLE\ar@{-}[ddlllllllllllll]\ar@{-}[ddlll]\ar@{-}[ddrrrrr]\ar@{-}[ddrrrrrrrrrrrrrrrrrrr]\ar@{-}[ddrrrrrrrrr]&&&& \CIRCLE\ar@{-}[ddlllllllllllllll]\ar@{-}[ddlllll]\ar@{-}[ddrrr]\ar@{-}[ddrrrrrrrrrrrrrrrrr]\ar@{-}[ddrrrrrrr]&&&& \CIRCLE\ar@{-}[ddlll]\ar@{-}[ddlllllll]\ar@{-}[ddlllllllllllllllll]\ar@{-}[ddrrrrrrrrrrrrrrr]\ar@{-}[ddrrrrr]&&&& \CIRCLE\ar@{-}[ddrrrrrrrrrrrrr]\ar@{-}[ddrrr]\ar@{-}[ddlllll]\ar@{-}[ddlllllllllllllllllll]\ar@{-}[ddlllllllll]&&&& \CIRCLE\ar@{-}[ddl]\ar@{-}[ddlll]\ar@{-}[ddlllll]\ar@{-}[ddlllllll]\ar@{-}[ddlllllllllllllllllllll]&&&& \CIRCLE\ar@{-}[ddlll]\ar@{-}[ddl]\ar@{-}[ddr]\ar@{-}[ddrrr]\ar@{-}[ddrrrrr]&&&&& \\
&&&&&&&&&&&&&&&&&&&&&&&&&&&&&&&&&&&&&&\\
\boxplus& \hspace{9pt}& \boxplus&\hspace{9pt}&
\boxplus&\hspace{9pt}& \boxplus&\hspace{9pt}& \boxplus&\hspace{9pt}&
\boxplus&\hspace{9pt}& \boxplus&\hspace{9pt}& \boxplus&\hspace{9pt}&
\boxplus&\hspace{9pt}& \boxplus&\hspace{9pt}& \boxplus&\hspace{9pt}&
\boxplus&\hspace{9pt}& \boxplus&\hspace{9pt}& \boxplus&\hspace{9pt}&
\boxplus&\hspace{9pt}& \boxplus&\hspace{9pt}& \boxplus&\hspace{9pt}&
\boxplus&\hspace{9pt}& \boxplus&\hspace{9pt}& \boxplus }$ }\\
\subfigure[Check nodes hidden] 
{
    \label{fig:ieee88:b}
$\xymatrix@M=0pt@W=0pt@R=20pt@C=20pt {
& \CIRCLE\ar@{-}[r]\ar@{-}[dl]\ar@{-}[ddl]\ar@{-}[drr]\ar@{-}[ddrr]& \CIRCLE \ar@{-}[dll]\ar@{-}[ddll]\ar@{-}[dr]\ar@{-}[ddr]&\\
\CIRCLE \ar@{-}[d]\ar@{-}[ddr]\ar@{-}[ddrr]&&& \CIRCLE\ar@{-}[d]\ar@{-}[ddl]\ar@{-}[ddll]\\
\CIRCLE\ar@{-}[dr]\ar@{-}[drr]&&& \CIRCLE\ar@{-}[dl]\ar@{-}[dll]\\
& \CIRCLE\ar@{-}[r]& \CIRCLE&\\
}$ } \caption{The topology of the dominant $(8,8)$ absorption set of the IEEE 802.3an code (not all
check node connections shown).}
\label{fig:ieee88}
\end{figure}
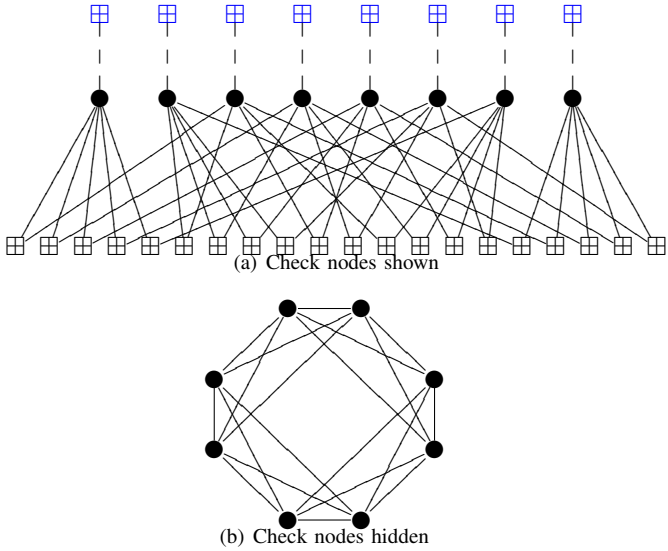

The subgraph induced by the dominant $(8,8)$ absorption set is
depicted in \figurename~\ref{fig:ieee88}. It can be seen that the
topology is highly symmetric and also looks like a codeword. From
\figurename~\ref{fig:ieee88:a} one sees that only one-sixth of the
neighboring check nodes will be unsatisfied when the set is in
error. In addition, the graph shown in
\figurename~\ref{fig:ieee88:b} is also full of cycles of several
lengths, similar to \figurename~\ref{fig15582:sub:b}.

The authors of \cite{DjuXuAbdLin03} have shown that the minimum
distance of this class of codes is lower bounded by $d_{\min}\geq
d_v+1$ and has to be even. So we have $d_{\min}\geq 8$ for this IEEE
802.3an LDPC code. As a corollary of our absorption set enumeration,
we tightened the lower bound to $d_{\min}>10$, because no $(a,0)$
absorption sets exist for $a\leq10$ as seen in
Table~\ref{table:8023ab} \cite{cs09}. In \cite{5454095}, it is shown
that there are at least $1,407$ weight-$14$ codewords, found by an
absorption set search algorithm. Therefore, the minimum distance is
narrowed down to $d_{\min}=12$ or $14$.\footnote{Evidence
suggests that $d_{\min}=14$. However our search program for $(12,0)$
absorption sets did not complete due to excessive time requirements.}

\subsection{Linear Algebraic Estimation of the Error Rate}\label{subsubsec1}

With the topologies and the multiplicities of the dominant
absorption sets of both codes in hand, we proceed to approximate
their contribution to the error rate.

\subsubsection{Tanner $[155, 64, 20], (3, 5)$ LDPC Code}

We apply the error formulation from \cite{itw09, cs09} to this
Tanner code using the $(8,2)$ absorption set structure.

The variable nodes in \figurename~\ref{fig15582:sub:a} are labeled
from $1$ to $8$. We also label the solid  edges from the variable
nodes by $1, 2, \dots, 22$ one by one from left to right. Also
denote the outgoing values from the variable nodes to the satisfied
check nodes along these edges by $x_i$, for example, $x_1,x_2,x_3$
leave variable node \ding{182}, $x_4,x_5, x_{6}$ variable node
\ding{183}, etc. Collect the $x_i$ in the length-$22$ column vector
$\mathbf{x}$, which is the vector of outgoing variable edge values
in the absorption set. Likewise, and analogously, let $\mathbf y$ be
the incoming edge values to the variable nodes, such that $y_j$
corresponds to the reverse-direction message on edge $j$,
$j=1,2,\dots,22$. Now, at iteration $i=0$,
\begin{equation}
\mathbf{x}_0 = \boldsymbol{\lambda},
\end{equation}
where the channel intrinsics vector $\boldsymbol{\lambda}$ is
defined in \eqref{mlambda}. It undergoes the following operation at
the check node:
\addtocounter{equation}{2}
\begin{equation}
\mathbf{y}_0 = \mathbf{C x}_0=\mathbf{C}\boldsymbol{\lambda},
\end{equation}
where $ \mathbf{C}$ is a permutation matrix that reflects the
incoming messages back to the absorption set. By induction, we
obtain at iteration $i=I$:
\begin{equation}\label{orgxI}
\mathbf{x}_I = \sum_{i=0}^I \left((\mathbf{V C})^i  \boldsymbol{\lambda} +
(\mathbf{V C})^{I-i}  \boldsymbol{\lambda}_i^{(\text{ex})}\right),
\end{equation}
where $\boldsymbol{\lambda}^{(\text{ex})}_i$ is the extrinsics
vector and defined in \eqref{mlambdaex}, and the variable node
addition matrix $\mathbf{V}$ is defined in \eqref{vmatrix}. Note
that $(\mathbf{V C})^0=\mathbf{I}$ and
$\boldsymbol{\lambda}_0^{(\text{ex})}=\mathbf{0}$.

{\setlength{\arraycolsep}{1.5pt}\renewcommand\arraystretch{0.4}
\begin{eqnarray}
\mathbf{V}&=&\begin{bmatrix}
0&1&1&&&&&&&&&&&&&&&&&&&\\
1&0&1&&&&&&&&&&&&&&&&&&&\\
1&1&0&&&&&&&&&&&&&&&&&&&\\
&&&0&1&1&&&&&&&&&&&&&&&&\\
&&&1&0&1&&&&&&&&&&&&&&&&\\
&&&1&1&0&&&&&&&&&&&&&&&&\\
&&&&&&0&1&1&&&&&&&&&&&&&\\
&&&&&&1&0&1&&&&&&&&&&&&&\\
&&&&&&1&1&0&&&&&&&&&&&&&\\
&&&&&&&&&0&1&&&&&&&&&&&\\
&&&&&&&&&1&0&&&&&&&&&&&\\
&&&&&&&&&&&0&1&&&&&&&&&\\
&&&&&&&&&&&1&0&&&&&&&&&\\
&&&&&&&&&&&&&0&1&1&&&&&&\\
&&&&&&&&&&&&&1&0&1&&&&&&\\
&&&&&&&&&&&&&1&1&0&&&&&&\\
&&&&&&&&&&&&&&&&0&1&1&&&\\
&&&&&&&&&&&&&&&&1&0&1&&&\\
&&&&&&&&&&&&&&&&1&1&0&&&\\
&&&&&&&&&&&&&&&&&&&0&1&1\\
&&&&&&&&&&&&&&&&&&&1&0&1\\
&&&&&&&&&&&&&&&&&&&1&1&0\\
\end{bmatrix}_{22\times 22}. \hspace{5mm}\label{vmatrix}
\end{eqnarray}}

Both $\mathbf{C}$ and $\mathbf{V}$ are square matrices and share the
dimension $ad_v-b$, which is the number of the solid edges in
\figurename~\ref{fig15582:sub:a}.

Now we calculate the maximum eigenvalue $\mu_{\max}$ of $\mathbf{V
C}$ and its corresponding unit-length eigenvector
$\mathbf{v}_{\max}$ in \eqref{umaxvmax}, since they are dominating
the magnitude of the power of $\mathbf{VC}$ in \eqref{orgxI}.
\begin{equation}\label{umaxvmax} \mu_{\max}\approx
1.7870, \quad \mathbf{v}_{\max}=\begin{bmatrix}
    v_1\\
    v_2\\
    v_3\\
    v_4\\
    v_5\\
    v_6\\
    v_7\\
    v_8\\
    v_9\\
    v_{10}\\
    v_{11}\\
    v_{12}\\
    v_{13}\\
    v_{14}\\
    v_{15}\\
    v_{16}\\
    v_{17}\\
    v_{18}\\
    v_{19}\\
    v_{20}\\
    v_{21}\\
    v_{22}\\
\end{bmatrix}\approx\begin{bmatrix}
    0.2369\\
    0.2369\\
    0.2273\\
    0.2031\\
    0.2031\\
    0.2651\\
    0.2254\\
    0.2254\\
    0.1660\\
    0.1261\\
    0.1483\\
    0.1483\\
    0.1261\\
    0.2031\\
    0.2651\\
    0.2031\\
    0.2369\\
    0.2369\\
    0.2273\\
    0.2201\\
    0.2201\\
    0.2544\\
\end{bmatrix}.
\end{equation}

The large value of $\mu_{\max}$ also underscores the dominance of this $(8,2)$ absorption set over
others listed in Table~\ref{table155as}.

Following \cite{itw09, cs09, allerton11}, the $(8,2)$ absorption set falls in error if
\begin{equation}\label{orgbeta}
\beta=\mathbf{x}_{I}^{\mathrm{T}}\cdot \mathbf{1} \leq 0.
\end{equation}

\begin{figure*}[!b]
\setcounter{MYtempeqncnt}{\value{equation}}
\setcounter{equation}{24} \vspace*{4pt} \hrulefill
\begin{eqnarray}
P_{\rm AS} &=& Q \left(\frac{\displaystyle A
m_\lambda\sum\limits_{i=0}^I \left(\frac{1}{\mu_{\max}^i}
\prod_{l=0}^{i}\frac{1}{g_l} \right)+B \sum\limits_{i=1}^I
\left(\frac{m_{\lambda^{(\text{ex})}}^{(i)}}{\mu_{\max}^i}
\prod_{l=1}^{i}\frac{1}{g_l} \right) }
    {\displaystyle \sqrt{ 2 C m_{\lambda} \left(\sum\limits_{i=0}^I \left(\frac{1}{\mu_{\max}^i} \prod_{l=0}^{i}\frac{1}{g_l} \right)\right)^2  + 2 D\sum\limits_{i=1}^I \left(\frac{m_{\lambda^{(\text{ex})}}^{(i)}}{\mu_{\max}^{2i}}   \left(\prod_{l=1}^{i}\frac{1}{g_l}\right)^2      \right)}  } \right)\label{orgprref}\\
P_{\rm AS} &=& Q \left(\frac{\displaystyle 2
m_\lambda\sum\limits_{i=0}^I \left(\frac{1}{\mu_{\max}^i}
\prod_{l=0}^{i}\frac{1}{g_l} \right)+2 \sum\limits_{i=1}^I
\left(\frac{m_{\lambda^{(\text{ex})}}^{(i)}}{\mu_{\max}^i}
\prod_{l=1}^{i}\frac{1}{g_l} \right) }
    {\displaystyle \sqrt{  m_{\lambda} \left(\sum\limits_{i=0}^I \left(\frac{1}{\mu_{\max}^i} \prod_{l=0}^{i}\frac{1}{g_l} \right)\right)^2  + \sum\limits_{i=1}^I \left(\frac{m_{\lambda^{(\text{ex})}}^{(i)}}{\mu_{\max}^{2i}}   \left(\prod_{l=1}^{i}\frac{1}{g_l}\right)^2      \right)}  } \right)\label{eq:ieeeformular}
\end{eqnarray}
\setcounter{equation}{\value{MYtempeqncnt}}
\end{figure*}
For simplicity, we separate this expression into two parts, namely
$\beta_1$ and $\beta_2$, as defined in \eqref{spectrual}.
\begin{equation}\label{spectrual}
\beta =\underbrace{\sum_{i=0}^I \left[(\mathbf{V C})^i
\boldsymbol{\lambda}\right]^{\mathrm{T}}\cdot\mathbf{1}}_{\triangleq\beta_1}
+ \underbrace{\sum_{i=0}^I \left[(\mathbf{V C})^{I-i}
\boldsymbol{\lambda}_i^{(\text{ex})}\right]^{\mathrm{T}}\cdot\mathbf{1}}_{\triangleq\beta_2}.
\end{equation}
Let {\setlength{\arraycolsep}{2pt}
\begin{eqnarray}
A&=& \sum_{j=1}^{22} v_j \label{eq:tanner155A}\\
B&=& v_{10}+v_{11}+v_{12}+v_{13}  \label{eq:tanner155B}\\
C&=& \left(v_1+v_2+v_3\right)^2+ \left(v_4+v_5+v_6\right)^2+ \left(v_7+v_8+v_9\right)^2 \nonumber \\
& & + \left(v_{10}+v_{11}\right)^2 + \left(v_{12}+v_{13}\right)^2 + \left(v_{14}+v_{15}+v_{16}\right)^2\nonumber\\
& & + \left(v_{17}+v_{18}+v_{19}\right)^2+ \left(v_{20}+v_{21}+v_{22}\right)^2 \label{eq:tanner155C}\\
D&=&
\left(v_{10}+v_{11}\right)^2+\left(v_{12}+v_{13}\right)^2\label{eq:tanner155D}
\end{eqnarray}}
and apply the spectral theorem,
\begin{equation}
 (\mathbf{V C})^i  \boldsymbol{\lambda}  \rightarrow \mu_{\max}^i
 \left( \boldsymbol{\lambda}^\mathrm{T}  \mathbf{v}_{\max} \right) \mathbf{v}_{\max},
\end{equation}
the means and the variances of $\beta_1$ and $\beta_2$,
individually, simplify to \eqref{eq:mbeta1}--\eqref{eq:varbeta2}.
\begin{eqnarray}
m_{\beta_1}&=&  A^2  \sum_{i=0}^I \frac{1}{\mu_{\max}^{i}} m_\lambda \label{eq:mbeta1}\\
\sigma_{\beta_1}^2&=&2 A^2 C \left(\sum_{i=0}^I \frac{1}{\mu_{\max}^{i}}\right)^2 m_\lambda \label{eq:varbeta1}\\
m_{\beta_2}&=& AB \sum_{i=1}^I \frac{m_{\lambda^{(\text{ex})}}^{(i)}}{\mu_{\max}^{i}} \label{eq:mbeta2}\\
\sigma_{\beta_2}^2&=&2 A^2 D  \sum_{i=1}^I \frac
{m_{\lambda^{(\text{ex})}}^{(i)}}{\mu_{\max}^{2i}}.\label{eq:varbeta2}
\end{eqnarray}

So, the probability of an absorption set falling in error is given by:
{\setlength{\arraycolsep}{1.5pt}\begin{eqnarray}
P_{\rm AS} &=& {\rm Pr} \left( \beta \leq 0 \right) \label{eq:orgformular}
= Q\left(\frac{m_{\beta_1}+m_{\beta_2}}{\sqrt{\sigma^2_{\beta_1}+\sigma^2_{\beta_2}}}\right)\\
&=& \! Q \left( \!\frac{\displaystyle A m_\lambda\sum\limits_{i=0}^I
\frac{1}{\mu_{\max}^i} +B \sum\limits_{i=1}^I
\frac{m_{\lambda^{(\text{ex})}}^{(i)}}{\mu_{\max}^i} }
    {\displaystyle \sqrt{ 2 C m_{\lambda} \left(\sum\limits_{i=0}^I \frac{1}{\mu_{\max}^i}\right)^2 \!\!\! + \! 2 D\sum\limits_{i=1}^I \frac{m_{\lambda^{(\text{ex})}}^{(i)}}{\mu_{\max}^{2i}}  } } \! \right).\hspace{5mm}  \label{orgpr}
\end{eqnarray}}

As illustrated, the factors $A, B, C$ and $D$ are determined by the absorption set topology. Knowledge
of $(a,b)$ and $d_v$ is encoded into them. For the Tanner code, $B$ or $D$ are much smaller than
$A$ or $C$, compared to the case of the dominant absorption set of the IEEE 802.3an code \cite{itw09, cs09}.
This implies that the critical extrinsic information has less impact on the error rate of the absorption set,
which makes it more troublesome than the $(8,8)$ absorption set of the IEEE 802.3an code.

Exchanging extrinsic values at the black check nodes in
\figurename~\ref{fig15582:sub:a} is only a first order approximation
of the actual messages returned, and the effect of the remaining $d_c-2$
inputs to these check nodes can be accounted for with an average,
iteration-depended gain factor which was computed in \cite{cs09} as
\begin{equation}\label{eq:correctionfactorg}
    g_l=\mathbb{E}\left[\prod_{i=1}^{d_c-2}\tanh\left(\frac{m_{\mu^{(\mathrm{ex})}}^{(i)}}{2}\right)\right],
\end{equation}
where $m_{\mu^{(\mathrm{ex})}}^{(i)}$ represents the mean of the
signals $\mu^{(\mathrm{ex})}$ from the variable to the check nodes,
and is computed using density evolution. With this refinement
the probability in \eqref{orgpr} is refined to \eqref{orgprref}.
(Note that $g_0$ is set to $1$.) This check node gain quickly grows to $1$ after
a few iterations, which implies that the external variable nodes have assumed
their corrected values with correspondingly large LLR values.

\subsubsection{IEEE 802.3an $[2048, 1723]$, $(6,32)$ LDPC Code}

The IEEE 802.3an LDPC code's dominant absorption set is shown in
\figurename~\ref{fig:ieee88}
and has eigenvalue and eigenvector given by \cite{cs09}:
\addtocounter{equation}{2}
\begin{eqnarray}
  \mu_{\max} &=& 4 \\
  \mathbf{v}_{\max} &=&
  \frac{1}{\sqrt{40}}\left[1,1,\dots,1\right]^{\mathrm{T}}_{40\times1}.\label{eq:8023vmax}
\end{eqnarray}

The symmetry is also reflected in the following coefficients.
{\setlength{\arraycolsep}{2pt}
\begin{eqnarray}
A&=& \sum_{j=1}^{40} v_j   = B\label{eq:ieeeA}\\
C&=& \left(v_1+v_2+v_3+v_4+v_5\right)^2 \nonumber\\
& &+ \left(v_6+v_7+v_8+v_{9}+v_{10}\right)^2 \nonumber\\
& &+ \left(v_{11}+v_{12}+v_{13}+v_{14}+v_{15}\right)^2 \nonumber\\
& &+ \left(v_{16}+v_{17}+v_{18}+v_{19}+v_{20}\right)^2 \nonumber\\
& &+\left(v_{21}+v_{22}+v_{23}+v_{24}+v_{25}\right)^2 \nonumber\\
& &+\left(v_{26}+v_{27}+v_{28}+v_{29}+v_{30}\right)^2 \nonumber\\
& &+\left(v_{31}+v_{32}+v_{33}+v_{34}+v_{35}\right)^2 \nonumber\\
& &+\left(v_{36}+v_{37}+v_{38}+v_{39}+v_{40}\right)^2 = D. \label{eq:ieeeC}
\end{eqnarray}}

Substituting \eqref{eq:ieeeA} and \eqref{eq:ieeeC} into
\eqref{orgprref}, the probability of the $(8,8)$ absorption set
falling in error is computed as \eqref{eq:ieeeformular}, which is
equivalent to setting $A=C=a=8$ and $B=D=b=8$.

\subsection{Error Probability Formula Refinement}

Besides the relative simplicity of \eqref{orgpr} and \eqref{orgprref},
one of the important insights we gain is that $\mu_{\max}$
and $\mathbf{v}_{\max}$ play a crucial role in the decoding failure mechanism. Absorption sets with large
$\mu_{\max}$ have more impact on the code performance, since they accelerate the growth of
erroneous LLRs within the set. The error formula can be strengthened, and in
this section, we revisit its derivation to obtain a more accurate formula.

\subsubsection{$P_{\mathrm{AS}}$}

We interpreted the probability of \eqref{orgbeta} happening as constituting the probability that an absorption set falls in error in \cite{itw09, cs09}. This condition can be strengthened because at the last iteration $I$, all elements of $\mathbf{x}_{I}$ are negative by virtue of \eqref{umaxvmax}. The failure probability is more accurately defined as
\begin{equation}\label{newxI}
P_{\mathrm{AS}}=\Pr(\mathbf{x}_{I} \leq \mathbf{0}).
\end{equation}
In addition, all elements of $\mathbf{x}_{I}$ are linear combinations of $\lambda_1, \lambda_2, \dots,
\lambda_a$ and $\lambda^{(\mathrm{ex})}_{i,1}, \lambda^{(\mathrm{ex})}_{i,2}, \dots,\lambda^{(\mathrm{ex})}_{i,b}$, where $i=1,2,\dots,I$. Therefore, \eqref{newxI} is equivalent to saying that
\begin{equation}\label{newxI2}
P_{\mathrm{AS}}=\Pr\left(\max_{j=1}^{ad_v-b}(x_{I,j}) \leq 0\right).
\end{equation}
In other words, if the message with the maximum value is negative at
the $I$-th iteration, then all the other variable nodes have
negative LLRs as well.

Analogously, we rewrite \eqref{orgxI} into two parts
$\boldsymbol{\beta}_1$ and $\boldsymbol{\beta}_2$, but as vectors
this time.
\begin{equation}
\boldsymbol{\beta}=\mathbf{x}_I =\underbrace{\sum_{i=0}^I (\mathbf{V
C})^i \boldsymbol{\lambda}}_{\triangleq\boldsymbol{\beta}_1} +
\underbrace{\sum_{i=0}^I (\mathbf{V C})^{I-i}
\boldsymbol{\lambda}_i^{(\text{ex})}}_{\triangleq\boldsymbol{\beta}_2}.
\end{equation}
Taking the mean of $\boldsymbol{\beta}_1$ and $\boldsymbol{\beta}_2$
we obtain {\setlength{\arraycolsep}{2pt}\begin{eqnarray}
\mathbf{m}_{\boldsymbol{\beta}_1} &=& m_{\lambda} \sum_{i=0}^I
(\mathbf{V
C})^i \mathbf{1},\label{eq:mean1vci}\\
\mathbf{m}_{\boldsymbol{\beta}_2} &=&
 \sum_{i=0}^{I-1} m^{(I-i)}_{\lambda^{(\mathrm{ex})}} (\mathbf{V C})^{i} \mathbf{1}\label{eq:mean2vci}
\end{eqnarray}}
for the $(8,8)$ absorption set of the IEEE 802.3an LDPC code, and
\begin{equation}\label{eq:mean2vci2}
\mathbf{m}_{\boldsymbol{\beta}_2} = \sum_{i=0}^{I-1}
m^{(I-i)}_{\lambda^{(\mathrm{ex})}} (\mathbf{V C})^{i} \big[
 0, \dots, 0, \underbrace{1, 1,}_{\text{node \ding{185}}} \underbrace{1, 1,}_{\text{node \ding{186}}} 0, \dots, 0 \big]^{\mathrm{T}}
\end{equation}
as mean of $\boldsymbol{\beta}_2$ for the $(8,2)$ absorption set of
the Tanner $[155,64,20]$ code.

It can be seen that the maximum entry of $\mathbf{v}_{\max}$ is dominating the elements of
$\mathbf{m}_{\boldsymbol{\beta}_1}+\mathbf{m}_{\boldsymbol{\beta}_2}$.

In the case of the $(8,2)$ absorption set of the Tanner code, $v_6$ or $v_{15}$ are the maximum entries
of $\mathbf{v}_{\max}$, as can be seen in \eqref{umaxvmax}.
Thus \eqref{newxI2} becomes
\begin{equation}\label{newxI3}
P_{\mathrm{AS}}=\Pr\left(\max_{j=1}^{22}(x_{I,j}) \leq 0\right) =\Pr\left(x_{I,6} \leq 0\right) .
\end{equation}
And our refined error probability formula can be derived
accordingly, using the same technique developed in Section
\ref{subsubsec1}. It will retain the form of \eqref{orgprref}, but is
more accurate.

For the IEEE 802.3an LDPC code, all $\mathbf{v}_{\max}$ entries have
equal size, as shown in \eqref{eq:8023vmax}. Selecting any $v_j$
for \eqref{newxI2} will result in \eqref{eq:ieeeformular} obtained
using \eqref{orgbeta} and \eqref{eq:orgformular}. Therefore, this
modification will not affect the result for the IEEE 802.3an code in
\cite{itw09,cs09}, and
\begin{equation}
\Pr\left(\mathbf{x}_{I}^{\mathrm{T}}\cdot \mathbf{1} \leq 0\right) =\Pr\left(\max_{j}^{40}(x_{I,j}) \leq 0\right).
\end{equation}

\subsubsection{Spectral Approximation}

In \eqref{orgprref} and \eqref{eq:ieeeformular}, we use the
approximation
\begin{equation}
 (\mathbf{V C})^i  \boldsymbol{\lambda}  \approx \mu_{\max}^i
 \left( \boldsymbol{\lambda}^\mathrm{T}  \mathbf{v}_{\max} \right) \mathbf{v}_{\max}, \quad i=1,2,\dots, I,
\end{equation}
to estimate the power of $\mathbf{V C}$. This is not very accurate
when $i$ is small. The results from the early iterations are of
great importance. So we will drop this approximation and use the
matrix formulation \eqref{eq:mean1vci}--\eqref{eq:mean2vci2} when
plotting error rate estimates.

\subsubsection{Numerical Results}

The refined formula predicting the error floor of the Tanner $[155,
64, 20], (3,5)$ code is plotted in
\figurename~\ref{fig:tannerformularis} as solid curves. The circles
are numerical results from importance sampling (IS) to support our
algebraic analysis. Error rates for the IEEE 802.3an code are shown in
\figurename~\ref{fig:ieeeformularis}.

\section{Error Floor Reduction}\label{boosting}

The error floor formula \eqref{orgprref} can be used to look for
contributing factors that come into play in the ultra-low BER
regime.

The absorption set structure determines the magnitude of
$\mu_{\max}$, which affects how fast the set LLRs grow. The coefficients
$A$, $B$, $C$ and $D$ depend on the code structure, as shown in
\eqref{eq:tanner155A}--\eqref{eq:tanner155D} and
\eqref{eq:ieeeA}--\eqref{eq:ieeeC}, and the topology of the absorption sets.
The mean of the intrinsics $m_{\lambda}$  dependents soley on the
channel LLR and has no impact on the error floor.

\begin{figure}[!t]
\centering {\setlength{\unitlength}{0.75mm}
\begin{picture}(125,75)
\put(-4,0){\includegraphics[scale=0.43]{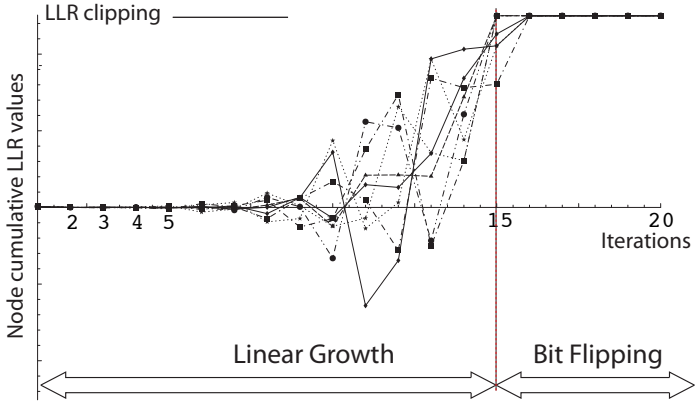}}
\end{picture} }
\caption{The accumulated LLRs at an $(8,8)$ absorption set nodes of the IEEE 802.3an LDPC code versus iterations, assuming all-zero codeword transmitted.}
\label{fig:ieeeTrapping}
\end{figure}

It is common understanding that simply increasing the maximum number
of iterations will not solve the trapping set problem.
This is because the absorption set will stabilize after a few iterations as soon
as the clipping levels are reached by the growing LLR messages. This convergence
to a bit-flipping operation is exemplified in \figurename~\ref{fig:ieeeTrapping}.
From Note~\ref{note1} now, the absorption set will remain in error even
for $I\rightarrow \infty$.

The decisive variable in the error floor formula is
$m_{\lambda^{(\mathrm{ex})}}$, the mean of the signals injected into
the absorption set through its unsatisfied check nodes, shown in
\figurename~\ref{fig15582:sub:a} and \figurename~\ref{fig:ieee88:a}.
It is evident from \eqref{orgprref} that the argument of the $Q$-function can
grow only if $m^{(i)}_{\lambda^{(\mathrm{ex})}}$ can outpace $\mu^{(i)}_{\max}$
as $i\rightarrow \infty$. Therefore, clipping of $m_{\lambda^{(\mathrm{ex})}}$ has
a major impact on the error probability.

By density evolution the extrinsics represented by
$m^{(i)}_{\lambda^{(\mathrm{ex})}}$ will grow to infinity, however, using
\begin{equation}
\label{eq:40}
\lambda_{j\to i}= 2\tanh^{-1}\left(\prod_{l\in
V_j\backslash\{i\}}\tanh\left(\frac{\lambda_{l \to j}}{2}\right)\right)
\end{equation}
in the decoding algorithm effectively clips
$\lambda_{i}^{(\mathrm{ex})}$ due to numerical
limitations of computing the inverse hyperbolic tangent
of a number close to unity.
This is observed and discussed in \cite{ButlerAllerton11} as saturation. Therefore,
instead of \eqref{eq:40}, we use the corrected min-sum algorithm \cite{mscf},
which computes the
check-to-variable LLR as
\begin{equation}\label{eq:mscf}
\lambda_{j\to i}=\left(\min_{l\in V_j\backslash\{i\}}
\left(\left|\lambda_{l\to j}\right|\right) +\mathrm{CT} \right) \prod_{l\in
V_j\backslash\{i\}}\text{sign}\left(\lambda_{l\to j}\right),
\end{equation}
where CT represents a correcting term dependent on the check node
degree,
\begin{equation}
\mathrm{CT}=\left\{\begin{array}{ll}
-\frac{\ln\left(d_c-1\right)}{4},  &  \text{if \ } \min_{l\in
V_j\backslash\{i\}}\left|\lambda_{l\to j}\right| \geq
\frac{3\ln\left(d_c-1\right)}{8}; \\
0, & \text{otherwise,}
\end{array}\right.
\end{equation}
to avoid the $\tanh(\cdot)$ calculation.

\begin{figure}[!t]
\centering
\includegraphics[scale=0.54]{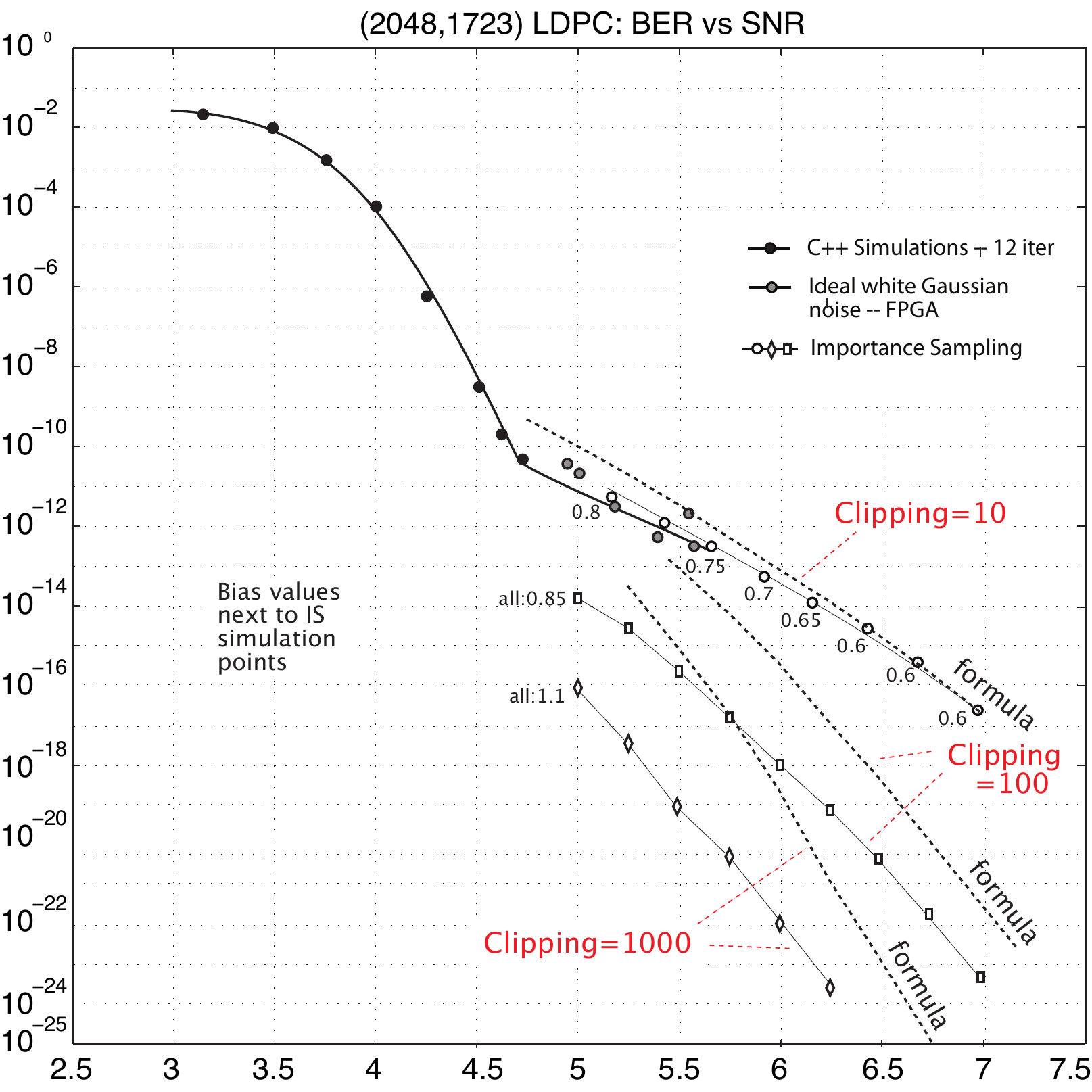}
\hspace*{\fill} $E_b/N_0$
\caption{Bit error rates of the IEEE 802.3an LDPC code using both formula
\eqref{eq:ieeeformular} and importance sampling (IS) with LLR clippings at 10,
100 and 1000, respectively. The iteration number is set to 10.}
\label{fig:ieeeformularis} 
\end{figure}

\figurename~\ref{fig:ieeeformularis} shows the error rates of the
IEEE 802.3an LDPC code in its error floor region. The dashed
curves plot \eqref{eq:ieeeformular} for $I=10$ and $m_\lambda$
and $m^{(i)}_{\lambda^{(\mathrm{ex})}}$ are bounded by $10$, which
is the clipping threshold. The circles are the numerical results of
importance sampling (IS) utilizing \eqref{eq:mscf} with the same $I$
and LLR clipping threshold. When the threshold is increased to
$100$, the error rate decreases as \eqref{eq:ieeeformular}
predicts, also shown in \figurename~\ref{fig:ieeeformularis}, and supported by IS
simulation. The error rate further decreases with the even bigger LLR
clipping value of 1,000.

Despite the high sensitivity of the formula to numerical issues, IS simulations
and analytical results agree to within a few tenths of a dB over the entire
range of $E_b/N_0$ values of interest.

\begin{figure}[!t]
\centering {\setlength{\unitlength}{0.75mm}
\begin{picture}(155,94)
\put(-5,-2){\includegraphics[scale=0.64]{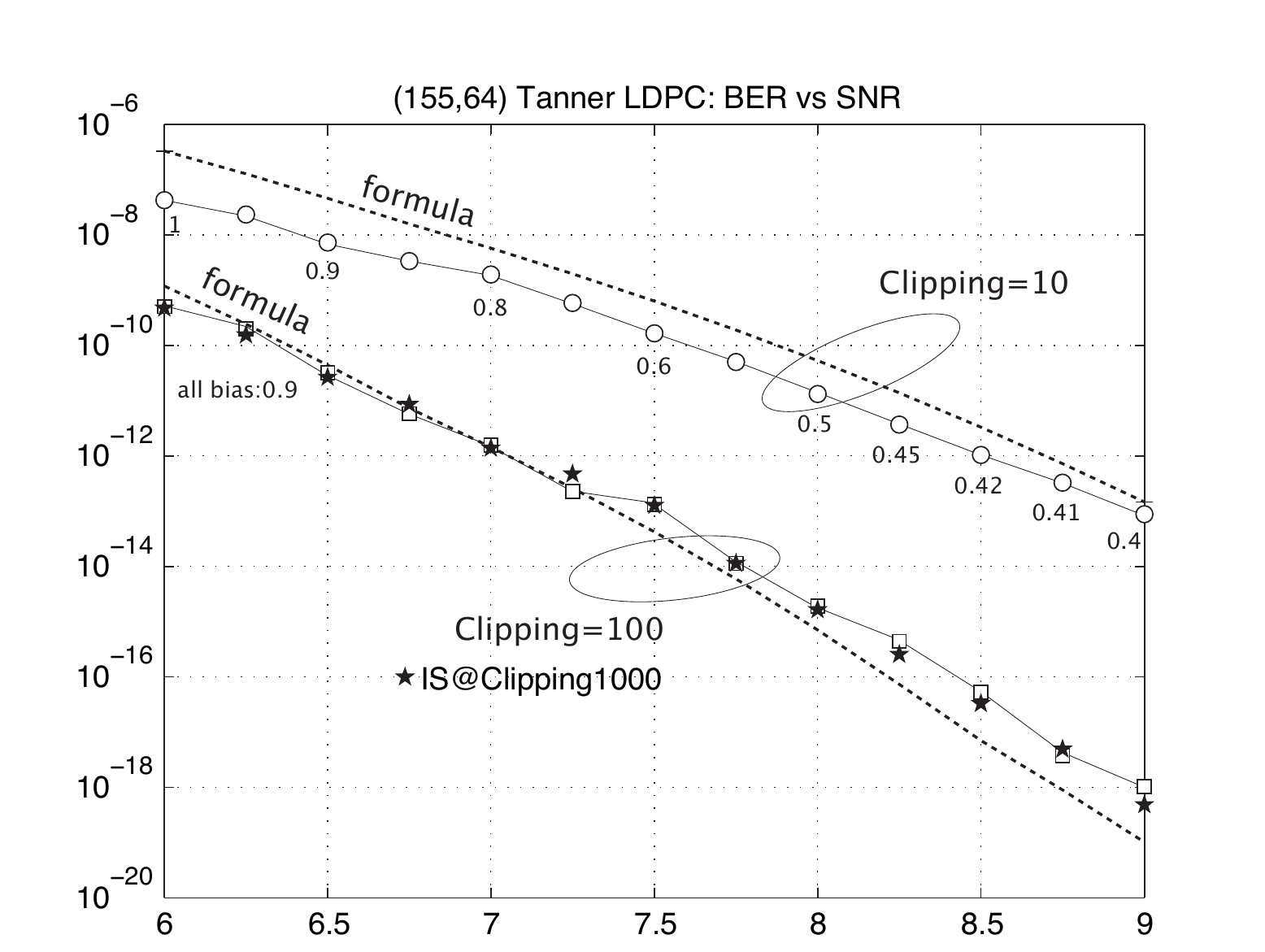}}
\end{picture}}
\hspace*{\fill} $E_b/N_0$
\caption{Bit error rates of the Tanner $[155, 64, 20]$ LDPC code using both formula \eqref{orgprref} and importance sampling (IS) with LLR clippings at 10, 100 and 1000, respectively. The maximum iteration
number is set to 50.}
\label{fig:tannerformularis} 
\end{figure}

Regarding the short code, \figurename~\ref{fig:tannerformularis} shows
the error rates of the Tanner $[155, 64, 20]$ LDPC code. Once again,
with higher LLR clipping value, \eqref{orgprref}
predicts that the error rate will decrease accordingly. This is
supported by IS when the clipping threshold is raised from 10 to
100. However, the IS results of LLR clipped at 1,000 are not as low
as suggested by \eqref{orgprref}, see the black stars in
\figurename~\ref{fig:tannerformularis}. This is due to the short
length of this code. Shortly after the decoding procedure begins, the
LLRs will become so correlated that the extrinsics
$\lambda_{i}^{(\mathrm{ex})}$ start to depart from the behavior predicted
by density evolution. However, eventually, the absorption set is still corrected
given more iterations are permitted \cite{allerton11}. The qualitative observations
made above are therefore valid also for short codes, even though the
quantitative prediction power of the error formulas breaks down.

\section{Iterations and Complexity}\label{secv}

Larger LLR clipping thresholds imply more complexity in practice, since
wider bit widths are needed to represent the messages.
\figurename~\ref{fig:ieeeformularis} shows how much gain can be
achieved by increasing the clipping values, in terms of lowering the
error rate of the IEEE 802.3an LDPC code. We note that the theoretical
results correlate well with the IS simulations.

Regarding the Tanner $[155, 64, 20]$ code, simply increasing the
clipping threshold to over 100 alone has no additional benefit, as
shown in \figurename~\ref{fig:tannerformularis}. The correlation
among the LLRs of this small code compromise the growth of the
extrinsic information entering the absorption sets.

Motivated by these observations, in this section we explore the impact of
message quantization and number of iterations on a code's error floor.
Arguably, the product of bit-width and number of iterations is an accurate measure
of the computational complexity of a decoder, since this number is directly
related to the switching activity of a digital decoder (see \cite{Schl2012}),
and hence also the energy expended by the decoder.

In order to verify our theoretical results, we resort again to importance sampling (IS).
IS is an ideal tool to explore variations of a decoder, such as finite precision operation,
where the impact of design changes need to be explored for ultra-low error rates.
Given the sensitivity of IS, we briefly discuss some major points
here and relate our own experiences with IS applied to LDPC decoding.

Importance sampling is a method where one increases the number of
significant events in low event-rate testing environments such as
bit error rate testing or simulation. The basic principle of
importance sampling is rooted in Monte-Carlo sampling. Specifically,
here we wish to evaluate the probability that noise carries
an original signal $x_0$ into the decision region of another signal,
thus causing a decoding error. If $N_s$ noise samples are selected
according to the channel's noise distribution, an estimate of this error
can be obtained as
\begin{equation}\label{mc_estimation}
\tilde{P}=\frac{1}{N_s}\sum_{i=1}^{N_s}w(y_i),
\end{equation}
which is an {\em unbiased estimate} of the true error
probability
\begin{equation}
P=\int_{\mathcal{D}}p(y|x_0)\; dy,
\end{equation}
where the weighting index is simply the error indicator
\begin{eqnarray}\label{ISweighting}
w(y)=\left\{
\begin{array}{ccl}
1,&& \text{if } y\in \mathcal{D};\\
0,&& \text{otherwise}.
\end{array}\right.
\end{eqnarray}
$\mathcal{D}$ is the signal space region where the decoder fails to
produce the correct output $x_0$, and $p(y|x_0)$ is the conditional
probability density function of the received signal $y$ given the
transmitted signal $x_0$, which is related to the noise
distribution, for example, the Gaussian noise density in our case.

Unfortunately, for low values of $P$, one has to generate on the
order of $10/P$ samples to obtain statistically relevant numbers
of error events. For instance, the error floor of the IEEE 802.3an
LDPC code appears below a bit error rate of $10^{-10}$, as shown in
\figurename~\ref{fig:ieeeformularis}, which requires $10^{11}$ to
$10^{12}$ samples to be simulated and tested, and with lowering
the error floor the number of test samples increases further.

One way of increasing the number of error events, or positive
counts, is to distort the noise distribution to cause more errors.
This is typically done by shifting the mean of the noise towards a
convenient boundary of $\mathcal{D}$ (mean-shift importance
sampling). The key questions are where to shift the transmitted signal
and by how much.

A priori knowledge of the dominant error mechanisms is extremely
important for proper use of IS, since otherwise a mean shift can
actually mask an error by moving the signal further away from the
dominant error event. Furthermore, the correct amount of the shift
is also important. If the shift value is too small, not enough simulation
speed up is achieved, if the shift value is too large, a phenomenon
called {\em over-biasing} causes the IS error estimate to dramatically
underestimate the true error contribution by the dominant event.
This happens when the biased simulated samples occur too far
away from the decision boundary, but inside the error region. These
samples are weighed with a index that is too small. Not enough
samples are generated close to the decision boundary from where the
majority of the actual error contribution originates.

Since absorption sets are examined as the primary causes of the
significant events in the error floor region, we add such a mean
shift, or \textit{bias}, towards the bits that make up the
absorption set. Having identified and enumerated the dominant
absorption sets of an LDPC code, we technically need to perform this
shift for each absorption set separately, but symmetries can often
be exploited in reducing this task. Gaussian distributed noise is
added to the thus biased codeword $x_0$. As a result, the  biased
received $y$ will have an increased chance of causing an error by
failing on the favored absorption set. Consequently the sample size
$N_s$ can be significantly reduced, which translates into what is
called ``the gain'' of importance sampling. Richardson \cite{Rich04}
used a version of IS where this mean shift assumes a continuous
distribution over which the simulations are averaged. This method
appears to alleviate the over-biasing, but also obscures the
phenomenon. In our approach we carefully choose the mean shift by
assuring that small changes do not alter the computed error rate.
Mean shift values used for some of our simulations are marked in the
figures. We finally wish to note that, unlike sometimes implied in
the literature, IS can only properly reproduce an error floor if the
causes of the error floor are sufficiently well understood to apply
proper biasing. Hence our two-step procedure in Section III.

Taking into account the bias value and the dominant absorption sets
selected in the operation, the estimation formula
\eqref{mc_estimation} along with the weighting factor
\eqref{ISweighting} must be adjusted accordingly, leading to a
weight term $w(y) \ll 1$. The combined effect of measuring more
significant events and ascribing them lower weight will produce the
same error rate measure in \eqref{mc_estimation} if the shifting is
done correctly. Figs.~\ref{fig:ieeeformularis} and
\ref{fig:tannerformularis} show the results of importance sampling
for the IEEE 802.3an and the Tanner $[155, 64, 20]$ LDPC codes
compared to our formulas for foating point calculations.

\begin{figure}[!t]
\centering
\includegraphics[scale=0.54]{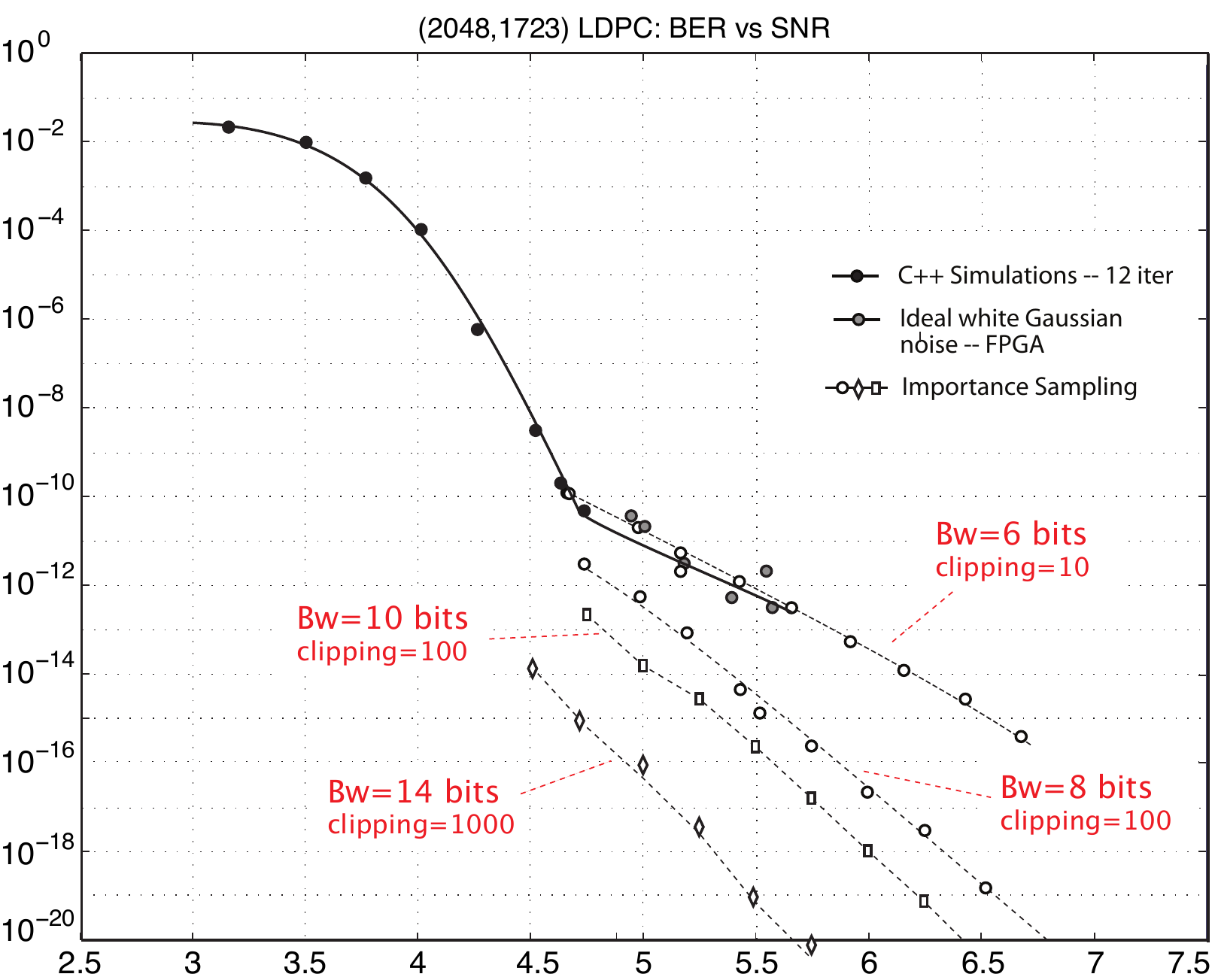}
\hspace*{\fill} $E_b/N_0$
\caption{Error rates of the IEEE 802.3an LDPC code via importance
sampling (IS) with finite precision, where the LLRs are clipped
at 10, 100 and 1000, respectively. The maximum iteration number is
preset at 10.}
\label{fig:ieeefiniteis} 
\end{figure}

In a hardware implementation, however, finite precision arithmetic is
used. The more digits used in the decoder, the more power and
computational effort is required, but better performance will results. It is
therefore vitally important for implementations to understand this cost-benefit
tradeoff. For the IEEE 802.3an code, IS simulations with both floating and fixed point
calculations at different LLR clipping values are shown in
\figurename~\ref{fig:ieeefiniteis}. Not surprisingly, for smaller clipping values,
smaller numbers of bits are required to adequately represent the messages.
While 6 bits of quantization are required for a clipping threshold of 10, 10 bits are
needed for a clipping threshold of 100, and 14 bits of quantization are required to
exploit the full benefit of a clipping threshold of~1000.

\section{Conclusion}\label{end}

We revisited the error floor of LDPC codes and showed
how the level of this floor can be controlled in a very large
range by a proper choice of message representation inside
the iterative message-passing decoder. In particular, the
maximum allowed value of these messages, the clipping
threshold, as well as the resolution of finite-precision arithmetic,
have a key impact on the level of the floor. For short codes,
additionally, very large iteration numbers may be required.

We verified our findings and error floor control mechanisms by
extending our recent linearized analysis theory, and numerically
verifying the results with importance sampling of the dominant
absorption sets for two representative
LDPC codes, namely the Tanner $[155,64,20]$ regular $(3,5)$
code and the larger IEEE 802.3an $[2048,1723]$ regular $(6,32)$
code. We conclude that the error floor can be lowered by many
orders of magnitude, or made to virtually disappear, with proper
setting of the message parameters, which allows the extrinsic
signals to outpace the error growth inside the absorption sets.


%





\ifCLASSOPTIONcaptionsoff
  \newpage
\fi



%



\bibliographystyle{IEEEtran}
\bibliography{IEEEabrv,ref}

\end{document}